\documentclass{article} 
\usepackage{graphicx}
\usepackage{lipsum}
\usepackage{euscript,amssymb}
\usepackage{epsfig}
\usepackage{color}

{\definecolor{RED}{rgb}{1,0,0}
\definecolor{GREEN}{rgb}{0,1,0}
\definecolor{BLUE}{rgb}{0,0,1}
\usepackage{amsfonts}

\usepackage{amsmath}

\usepackage{amssymb}

\usepackage{fancyhdr}
\def\nn{\nonumber\\}
\def\be{\begin{equation}}
\def\ee{\end{equation}}
\def\bea{\begin{eqnarray}}
\def\eea{\end{eqnarray}}

\def\bwt{\begin{widetext}}
\def\ewt{\end{widetext}}

\usepackage{esint}
\usepackage[unicode=true, pdfusetitle,
 bookmarks=true,bookmarksnumbered=false,bookmarksopen=false,
 breaklinks=false,pdfborder={0 0 1},backref=false,colorlinks=false]
 {hyperref}

\newcommand{\f}[2]{\frac{#1}{#2}}
\begin{document}

\title{Collapse and dispersal of a homogeneous spin fluid in Einstein-Cartan theory}
\author{M. Hashemi,\, S. Jalalzadeh \footnote{s-jalalzadeh@sbu.ac.ir} \,and\, A. H. Ziaie
\\\\Department of Physics, Shahid Beheshti University,\\ G. C., Evin,Tehran, 19839, Iran}
\maketitle



\begin{abstract}
In the present work, we revisit the process of gravitational collapse of a spherically symmetric homogeneous dust fluid which is known as the Oppenheimer-Snyder (OS) model \cite{OS}. We show that such a scenario would not end in a spacetime singularity when the spin degrees of freedom of fermionic particles within the collapsing cloud are taken into account. To this purpose, we take the matter content of the stellar object as a homogeneous Weyssenhoff fluid which is a generalization of perfect fluid in general relativity (GR) to include the spin of matter. Employing the homogeneous and isotropic FLRW metric for the interior spacetime setup, it is shown that the spin of matter, in the context of a negative pressure, acts against the pull of gravity and decelerates the dynamical evolution of the collapse in its later stages. Our results bode a picture of gravitational collapse in which the collapse process halts at a finite radius whose value depends on the initial configuration. We thus show that the spacetime singularity that occurs in the OS model is replaced by a non-singular bounce beyond which the collapsing cloud re-expands to infinity. Depending on the model parameters, one can find a minimum value for the boundary of the collapsing cloud or correspondingly a threshold value for the mass content below which the horizon formation can be avoided. Our results are supported by a thorough numerical analysis.





\end{abstract}



\section{Introduction}
\label{Intro}
One of the most important questions in a gravitational theory (such as GR) and relativistic astrophysics is the gravitational
collapse of a massive star under its own gravity at the end of its life cycle. A process in which a sufficiently massive
star undergoes a continual gravitational collapse on exhausting its nuclear fuel, without achieving an equilibrium state \cite{joshb}. According to the singularity theorems in GR \cite{sintheo}, the spacetimes describing the solutions of the Einstein\rq{}s field equations in a typical collapse scenario would inevitably admit singularities\footnote{These are the spacetime events where the metric tensor is undefined or is not suitably differentiable, the curvature scalars and densities are infinite and the existing physical framework would break down \cite{hawsin}.}. These theorems are based on three main assumptions under which the existence of a spacetime singularity is foretold in the form of geodesic incompleteness in the spacetime. The first premise is in the form of a suitable causality condition that ensures a physically reasonable global structure of the spacetime. The second premise is an energy condition that requires the positivity of the energy density at the classical regime as seen by a local observer. The third one demands that gravity be so strong in the sense that trapped surface\footnote{A trapped surface is a closed two-surface on which
both in-going as well as out-going light signals normal to it are necessarily converging \cite{FROLOV}.} formation must occur during the dynamical evolution of a continual gravitational collapse.
\par   The first detailed treatment of the gravitational collapse of a massive star, within the framework of GR, was published by Oppenheimer and Snyder \cite{OS}. They concluded that gravitational collapse of a spherically symmetric homogeneous dust cloud would end in a black hole. Such a black hole is described by the presence of a horizon which covers the spacetime singularity. This scenario provides the basic motivation for the physics of black holes, and the cosmic censorship conjecture (CCC) \cite{CCC}. This conjecture sates that, the spacetime singularities that develop in a gravitational collapse scenario are necessarily covered by the event horizons, thus ensuring that the collapse end-product is a black hole only. As no proof, or an stringent mathematical formulation of the CCC has been available so far, a great deal of effort has been made in the past decades to perform a detailed study of 
several collapse settings in GR, in order to extend our understanding from the vague corners of this phenomenon.
\par
While black hole physics has given rise to several interesting theoretical as well as astrophysical progresses, it is necessary, however, to investigate more realistic collapse settings in order to put black hole physics on a firm status. This is because the OS model is rather idealized and pressures as well as inhomogeneities within the matter distributions, would play an important role in the collapse dynamics of any realistic stellar object. It is therefore of significant importance to broaden the study of gravitational collapse to more realistic models in order to deal with this question: what 
ways there are the possible departures in final outcomes, as opposed to the homogeneous dust cloud collapse? Within this context, several gravitational collapse settings have been investigated over the past years which represent the occurrence of naked singularities\footnote{In this case, the horizons are delayed or failed to form during collapse process, as governed by the internal dynamics of the collapsing object. Then, the scenario where the super-dense regions are visible to external observers occurs, and a visible naked singularity forms \cite{joshb}.}. Work along this line has been reported in the literature within a variety of models; among them we quote the role of inhomogeneities within the matter distribution on the final fate of gravitational collapse \cite{Joshi-inhom}, collapse of a perfect fluid with heat conduction \cite {heat-con}, effects of shear on the collapse end-product \cite{shear},  and collapse process in the context of different gravitational theories \cite{alt-col} (see also \cite{REC} for recent reviews).
\par
On the other hand, though GR has emerged as a highly successful theory of gravitation, it suffers from the occurrence of spacetime singularities under physically reasonable conditions. It is therefore plausible to seek for the alternative theories of gravitation whose geometrical attributes are not present in GR. This allows for the inclusion of more realistic matter fields within the structure of stellar objects, in order to cure the singularity problem. In this regard, since the realistic stars are made up fermions, it would be difficult to reject the role of intrinsic angular momentum (spin) of fermions in collapse studies. As we shall see, the inclusion of spin of fermions and thus its possible effects on the collapse dynamics could be of significant importance specially at the late stages of the collapse setting where theses effects could go against the gravitational attraction to ultimately balance it. In such a scenario the collapse may no longer terminate in a spacetime singularity and instead is replaced by a bounce, a point at which the contraction of matter cloud stops and an expanding phase begins. However, if the spin effects are explicitly present, then GR will no longer be the relevant theory to describe the collapse dynamics. In GR, the energy-momentum tensor couples to the metric, while in the presence of fermions, it is expected that the intrinsic angular momentum is coupled to a geometrical quantity related to the rotational degrees of freedom in the spacetime, the so called spacetime torsion. This obviously is not possible in the ambit of GR so that one is necessitated to modify the theory in order to introduce torsion and relate it to the spin degrees of freedom of fermions. This point of view suggests a spacetime manifold which is non-Riemannian. One such framework, within which the inclusion of spin effects of fermions can be worked out and thus will allow non-trivial dynamical consequences to be extracted is the Einstein-Cartan (EC) theory \cite{ECT,ECT1,ECT2,ECT3}. Within this context, many cosmological models have been found in which the unphysical big bang singularity is replaced with a bounce at a finite amount  the scalar factor \cite{spin-bounce,POPPRD}. From another perspective, the research of the recent years has shown that in the final stages of a typical collapse scenario where a high energy regime governs, the effects of quantum gravity would regularize the singularity that happens in the classical model \cite{Bojowald-PRL-2005}. In cosmological settings, it is shown that non-perturbative quantum geometric effects in loop quantum cosmology would replace the classical singularity by a quantum bounce in the high energy regime where the loop quantum modifications are dominant \cite{SVV}. However, since the full quantum theory of gravity has not yet been discovered, investigating the repulsive spin effects of fermions, which is more physically reasonable and confirmed observationally, on the final state of collapse could be well-motivated. The organization of this paper is as follows: In Sec. \ref{EC} we give a brief review on the field equations in EC theory and the phenomenological Weyssenhoff model. In Sec. \ref{SSR}, we study the collapse dynamics in the presence of spin effects  and the possibility of singularity removal. Finally, Conclusions are drawn in Sec. \ref{con}.
\section{Einstein-Cartan theory}\label{EC}
As we know, the dynamics of the gravitational field, i.e., the metric field, in GR is described by the Hilbert-Einstein action with the Lagrangian which is linear in curvature scalar. Contrary to GR, in the gravity with torsion there is a considerable freedom in constructing the dynamical 
scheme, since one can define much more invariants from torsion and curvature tensors. There are two most attractable classes of models, namely: the EC theory and the quadratic theories. In this work we are interested in the former one for which the action integral is given by 
\be S=\int  d^4x\sqrt{-g}
\left\{\f{-\hat{R}}{2\kappa}+\mathcal{L}_m\right\}, \label{action}
\ee
where $\kappa=8\pi G$ (we set $c=1$) is the gravitational coupling constant, $\mathcal L_m$ is the Lagrangian for the material fields and $\hat{R}$ is the EC curvature scalar constructed out of the general asymmetric connection $\hat{\Gamma}^{\alpha}_{~\mu\nu}$, i.e., the connection of Riemann-Cartan manifold. 
The torsion tensor $T^{\alpha}_{~\mu\nu}$ is defined as the antisymmetric part of the affine connection, given by
\be\label{TT}
T^{\mu}_{~\alpha\beta}=\f{1}{2}\left[\hat{\Gamma}^{\mu}_{~\alpha\beta}-\hat{\Gamma}^{\mu}_{~\beta\alpha}\right].
\ee
From the metricity condition, $\hat{\nabla}_{\alpha}g_{\mu\nu}=0$ we can find the affine connection as
\be\label{AFC}
\hat{\Gamma}^{\mu}_{~\alpha\beta}=\big\{^{\,\mu} _{\alpha\beta}\big\}+K^{\mu}_{~\alpha\beta},
\ee
where the first part being the Christoffel symbols and the second part being the contorsion tensor defined as 
\be\label{contortion}
K^{\mu}_{~\alpha\beta}=T^{\mu}_{~\alpha\beta}+T_{\alpha\beta}^{~~\,\mu}+
T_{\beta\alpha}^{~~\,\mu}.
\ee
Extremizing the total action with respect to contorsion tensor gives the Cartan field equation as
\be\label{FEEC}
T^{\alpha}_{~\mu\beta}-\delta^{\alpha}_{\,\beta}T^{\gamma}_{~\,\mu\gamma}+\delta^{\alpha}_{\,\mu}T^{\gamma}_{~\,\beta\gamma}=-\f{1}{2}\kappa\tau_{\mu\beta}^{~~\alpha},
\ee
where $\tau^{\mu\alpha\beta}=2\left(\delta\mathcal L_m/\delta K_{\mu\alpha\beta}\right)/\sqrt{-g}$ is the spin angular momentum tensor \cite{ECT3}. It is worth noting that the equation governing the torsion tensor is an equation of pure algebraic type, i.e., the torsion is not allowed to propagate beyond the matter distribution as a torsion wave or through any interaction of non-vanishing range \cite{ECT3}; therefore it can be only nonzero inside material bodies. Varying the total action with respect to the metric tensor leads to the Einstein\rq{}s field equation with additional terms on the curvature side, that are quadratic in the torsion tensor \cite{Venzo}. Substituting for the torsion tensor, from equation (\ref{FEEC}), into these terms we get the combined field equations as \cite{ECT3,POPPRD,NPOP,Venzo}
\bea\label{COMFE}
G_{\mu\beta}\left(\{\}\right)=\kappa\left({\mathcal T}_{\mu\beta}+\Sigma_{\mu\beta}\right),
\eea
where ${\mathcal T}_{\mu\beta}=2\left(\delta\mathcal L_m/\delta g^{\mu\beta}\right)/\sqrt{-g}$, is the dynamical energy-momentum tensor and $\Sigma_{\mu\beta}$ can be considered as representing the contribution of an effective spin-spin interaction \cite{ECT3}, i.e., the product terms
\bea\label{SSCI}
\Sigma_{\mu\beta}&=&\f{1}{2}\kappa\bigg[\tau_{\mu\alpha}^{~~~\alpha}\tau_{\beta\gamma}^{~~~\!\!\gamma}-\tau_{\mu}^{~\alpha\gamma}\tau_{\beta\gamma\alpha}-\tau_{\mu}^{~\alpha\gamma}\tau_{\beta\alpha\gamma}\nn&+&\f{1}{2}\tau^{\alpha\gamma}_{~~~\mu}\tau_{\alpha\gamma\beta}+\f{1}{4}g_{\mu\beta}\left(2\tau_{\alpha\gamma\epsilon}\tau^{\alpha\epsilon\gamma}
-2\tau_{\alpha~\gamma}^{~\gamma}\tau^{\alpha\epsilon}_{~~~\epsilon}
+\tau^{\alpha\gamma\epsilon}\tau_{\alpha\gamma\epsilon}\right)\bigg].
\eea
\par
It now is obvious that the second term on the right hand side of (\ref{COMFE}), represents a correction (though very weakly at ordinary densities as this term carries a factor $\kappa^2$) to the dynamical energy-momentum tensor, which takes into account the spin contributions to the geometry of the manifold\footnote{We note that if the spin is switched off, the field equation (\ref{COMFE}) reduces to the ordinary Einstein\rq{}s field equation.}. However, the spin corrections are significant only at the late stages of gravitational collapse of a compact object where super-dense regions of extreme gravity are involved. Therefore, there is  a good motivation to investigate the collapse process of material fluid sources which are endowed with spin. Let us now apply equation (\ref{COMFE}) to estimate the influence of spin in the case of Weyssenhoff fluid which generalizes the perfect fluid of GR to the case of non-vanishing spin. This model of the fluid was first studied by Weyssenhoff and Raabe \cite{W1947} and extended by Obukhov and Korotky in order to build cosmological models based on the EC theory \cite{KCQG1987}. In the model presented in this paper we employ an ideal Weyssenhoff fluid which is considered as a continuous medium whose elements are characterized by the intrinsic angular momentum (spin) of particles. In this  model  the spin density is described by the second-rank antisymmetric tensor $S_{\mu\nu}=-S_{\nu\mu}$. The spin tensor for the Weyssenhoff fluid is then postulated to be 
\be\label{FC}
\tau_{\mu\nu}^{~~\alpha}=S_{\mu\nu}{\rm U}^{\alpha},
\ee
where ${\rm U}^{\alpha}$ is the four-velocity of the fluid element. The Frenkel condition which arises by varying the Lagrangian of the sources \cite{KCQG1987} requires $S^{\mu\nu}{\rm U}_{\nu}=0$. This condition further restricts the torsion tensor to be traceless.
From the microscopical viewpoint, a randomly oriented gas of fermions is the source for the spacetime torsion. However, we have to treat this issue from a macroscopic perspective, that means we need to perform suitable spacetime averaging. In this respect, the average of the spin density tensor vanishes, $\langle S_{\mu\nu} \rangle=0$ \cite{ECT3,G1986}. But even with vanishing this term at macroscopic level, the square of spin density tensor $S^2=\f{1}{2}\langle S_{\mu\nu}S^{\mu\nu}\rangle$ contributes to the total energy-momentum tensor. Taking these considerations into account, the relations (\ref{COMFE})-(\ref{FC}) then give the Einstein\rq{}s field equation with spin correction terms \cite{ECT3,NPOP}
\be\label{EFESSP}
G_{\mu\nu}=\kappa\left(\rho+p-\f{\kappa}{2}S^2\right){\rm U}_{\mu}{\rm U}_{\nu}-\kappa\left(p-\f{\kappa}{4}S^2\right)g_{\mu\nu},
\ee
where $\rho$ and $p$ are the usual energy density and pressure of the perfect fluid satisfying a barotropic equation of state $p=w\rho$.  
Thus, the EC equation for such a spin fluid would be equivalent to the Einstein\rq{}s equation for a perfect fluid with the effective energy density $\rho_{\rm eff}=\rho-\f{\kappa}{4}S^2$ and effective pressure $p_{\rm eff}=p-\f{\kappa}{4}S^2$. This estimate shows that, the contribution of spin of fermions to the gravitational interaction is negligible in the case of normal matter densities (e.g., in the early stages of the collapse process) while in the late stages of the collapse where one encounters the ultra-high energy densities, it is the spin contribution that decides the final fate of the collapse scenario. This is the subject of our next discussion.


\section{Spin effects on the collapse dynamics and singularity removal}
\label{SSR}
The study of gravitational collapse of a compact object and its importance in relativistic astrophysics was initiated since the work of Datt \cite{Datt-ZPhys} and OS \cite{OS} (see also \cite{PEDFAB} for a pedagogical discussion). This model which simplifies the complexity of such an astrophysical scenario describes the process of gravitational collapse of a homogeneous dust cloud with no rotation and internal stresses in the framework of GR. Assuming the interior geometry of the collapsing object to be that of FLRW metric, they investigated the dynamics of the continual gravitational collapse of such a matter distribution under its own weight and showed that for an observer comoving with the fluid, the radius of the star crushes to zero size and the energy density diverges in a finite proper time. For this idealized model which showed that a black hole is developed as the collapse end-state, the only evolving portion of the spacetime is the interior of the collapsing object while the exterior spacetime remains as that of Schwarzschild solution with a dynamical boundary. However, in more realistic scenarios, the dynamical evolution of a collapse setting would be significantly different in the later stages of the collapse where the inhomogeneities are introduced within the densities and pressures. These effects could alter the dynamics of the horizons and consequently, the fate of collapse scenario \cite{PSMASAR}. 
\par
The spin effects within more realistic stellar collapse models could have considerable effects on the collapse dynamics as we shall see in this section. In order to deal with this purpose, the matter content of the collapsing object is taken as a homogeneous and isotropic Weyssenhoff fluid that collapses under its own gravity. We then parametrize the interior line element as
\be\label{FRWL}
ds^2=dt^2-\f{a(t)^2dr^2}{1-kr^2}-R^2(t,r)d\Omega^2,
\ee
where $R(t,r)=ra(t)$ is the physical radius of the collapsing star with $a(t)$ is the scale factor, $k$ is a constant that is related to the curvature of spatial metric and $d\Omega^2$ is the standard line element on the unit two-sphere. The field equations then read 
\begin{align}\label{FE12}
\left\{\begin{array}{l}
\left(\f{\dot{a}}{a}\right)^2+\f{k}{a^2}=\f{8\pi G}{3}\rho-\f{(4\pi G)^2}{3}S^2,
                      \\
                      \\
                      \\
\f{\ddot{a}}{a}=-\f{4\pi G}{3}(\rho+3p)+\f{2}{3}(4\pi G)^2S^2.
                   \end{array}\right.
\end{align}
The contracted Bianchi identities give rise to the continuity equation as
\be\label{ConEq}
\dot{\rho}_{\rm eff}=-3\f{\dot{a}}{a}(\rho_{\rm eff}+p_{\rm eff}),
\ee
whence we have 
\be\label{KAS}
\dot{\rho}=-3\f{\dot{a}}{a}(\rho+p),~~~(S^2\dot{)}=-6\f{\dot{a}}{a}S^2.
\ee
The first part of the above equations give
\be\label{RHOA}
\rho=\rho_i \left(\frac{a}{a_i}\right)^{-3(1+w)},
\ee
where $\rho_i$ is the initial energy density profile and $a_i$ is the initial value of the scale factor at the initial epoch. A suitable averaging procedure leads to the following relation between the spin squared and energy densities as \cite{NPPL1983}
\be\label{S2}
S^2=  \f{\hbar^2}{8}A_w^{-\f{2}{1+w}}\rho^{\f{2}{1+w}},
\ee
where $A_w$ is a dimensional constant that depends on the equation of state parameter. It should be noticed that substituting (\ref{RHOA}) into the above expression leads to $S^2 \propto a^{-6}$ which is nothing but the solution of the second part of (\ref{KAS}). The field equations can then be re-written as
\begin{align}\label{FERE}
\left\{\begin{array}{l}
\left(\f{\dot{a}}{a}\right)^2+\f{k}{a^2}=2Ca^{-3(1+w)}-Da^{-6},
                      \\
                      \\
                      \\
\f{\ddot{a}}{a}=-C(1+3w)a^{-3(1+w)}+2Da^{-6},
                   \end{array}\right.
\end{align}
where $ C=\f{4\pi G}{3}\rho_i {a_i}^{3(1+w)}$ and $D=\f{(4\pi G)^2}{24}\hbar^2 A_w^{-\f{2}{1+w}}\rho_i^{\f{2}{1+w}}a_i^6$. Next we proceed to study the collapse evolution for different values of the spatial curvature. We assume that the star begins its contraction phase from a stable situation, i.e. $\dot{a}(t_i)=0$, where $t_i$ is the initial time at which the collapse commences. Thus, from the first part of (\ref{FERE}) we find
\be\label{k}
k=\left[\f{2C}{D}-a_i^{3(w-1)}\right]Da_i^{-(1+3w)}.
\ee
Depending on the sign of the expression in double brackets, the constant $k$ may be either positive, negative or zero. Therefore we may write
\begin{align}\label{signk}
\left\{\begin{array}{l}
k>0~~~~~~~~\f{2C}{D}>a_i^{3(w-1)},
                      \\
                      \\
k\leq0~~~~~~~~\f{2C}{D}\leq a_i^{3(w-1)}.
\end{array}\right.
\end{align}

Let us consider the dust fluid ($w=0$) for which the solution of ($k=0$) clearly represents an expanding solution. For the case ($k<0$) the collapse velocity is non-real which is physically implausible \cite{SAS-PTP-88}. Thus the only remained case is $k>0$ for which we are to investigate the collapse dynamics for large and small values of the scale factor, i.e., the early and late stages of the collapse process, respectively.

In the early stages of the collapse, the spin contribution is negligible and thus the first part of (\ref{FERE}) can be approximated as
\be\label{FEAP}
\dot{a}^2\cong-k+\f{2C}{a}.
\ee
Performing the transformation $ad\xi=\sqrt{k}dt$ we get the solution as
\begin{align}\label{kpl}
\left\{\begin{array}{l}
a(\xi)=\f{C}{k}(1\pm \cos(\xi)),
                      \\
                      \\
t(\xi)=\f{C}{k^{\f{3}{2}}}(\xi \pm \sin(\xi))+t_i,
                      \\
\end{array}\right.
\end{align}
where according to equation (\ref{FEAP}) $a_i\cong2C/k$.

For the collapse evolution where the scale factor has become small enough and the spin effects are dominant, the $k$ term in the first Friemann equation (\ref{FERE}) can be neglected compared to the rest. We then get
\be\label{FERTorres}
\dot{a}^2\cong\f{2C}{a}-\f{D}{a^4},
\ee
for which the solution is given by
\be\label{kps}
a(t)=\left\{a_s^3+\f{9}{2}C(t-t_s)^2-\sqrt{18C}(t-t_s)\sqrt{a_s^3-\f{D}{2C}}\right\}^{\f{1}{3}},
\ee
where $t_s>t_i$ represents the time at which the small scale factor regime starts at a finite value of the scale factor $a_s<a_i$.  The solution (\ref{kps}) exhibits a bounce at a finite time, say $t=t_b$, where the collapse halts ($\dot{a}(t_b)=0$) at a minimum value of the scale factor given by
\bea\label{amin}
a_{{\rm min}}=\left[\f{D}{2C}\right]^{\f{1}{3}}=\left[\f{\pi G\hbar^2\rho_i}{4A_0^2}\right]^{\f{1}{3}}a_i.
\eea
We note that for $a>a_{{\rm min}}$, equation (\ref{kps}) is always real. 

For a physically reasonable collapse setting the weak energy condition (WEC) must be satisfied. This condition states that for any non-spacelike vector field $T_{\alpha\beta}^{\rm{eff}}V^\alpha V^{\beta}\geq0$, which for our model amounts to $\rho_{\rm{eff}}\geq0$ and $\rho_{\rm{eff}}+p_{\rm{eff}}\geq0$. The first inequality suggests that $\rho\geq2\pi G S^2$ while the latter implies $\rho\geq4\pi G S^2$. The first inequality with the use of (\ref{S2}) gives 
\be\label{WEC}
\rho\leq\f{4A_0^2}{\pi G \hbar^2}=\rho_i\left(\f{a_i}{a_{\rm{min}}}\right)^3,
\ee
whereby considering (\ref{RHOA}) we arrive at $a/a_{\rm{min}}\geq1$. Since the scale factor never reaches the values smaller than $a_{\rm{min}}$, this inequality is always held implying the satisfaction of positive energy density condition. Moreover, the second inequality with similar calculations for dust tells us  $a/a_{\rm{min}}\geq 2^{\f{1}{3}}$. This means that in the later stages of the collapse as governed by a spin dominated regime, WEC is violated. Such a violation of the weak energy condition can be compared to the models where the quantum effects in the collapse scenario have been discussed \cite{Daniele-PRD-2013}. In brief, we have WEC violation for the following interval
\be\label{WECC}
a_{{\rm min}}\le a < 2^{\f{1}{3}}a_{{\rm min}}.
\ee
\subsection{Numerical analysis}
\label{NA}
\begin{figure}
\begin{center}
\includegraphics[scale=0.335]{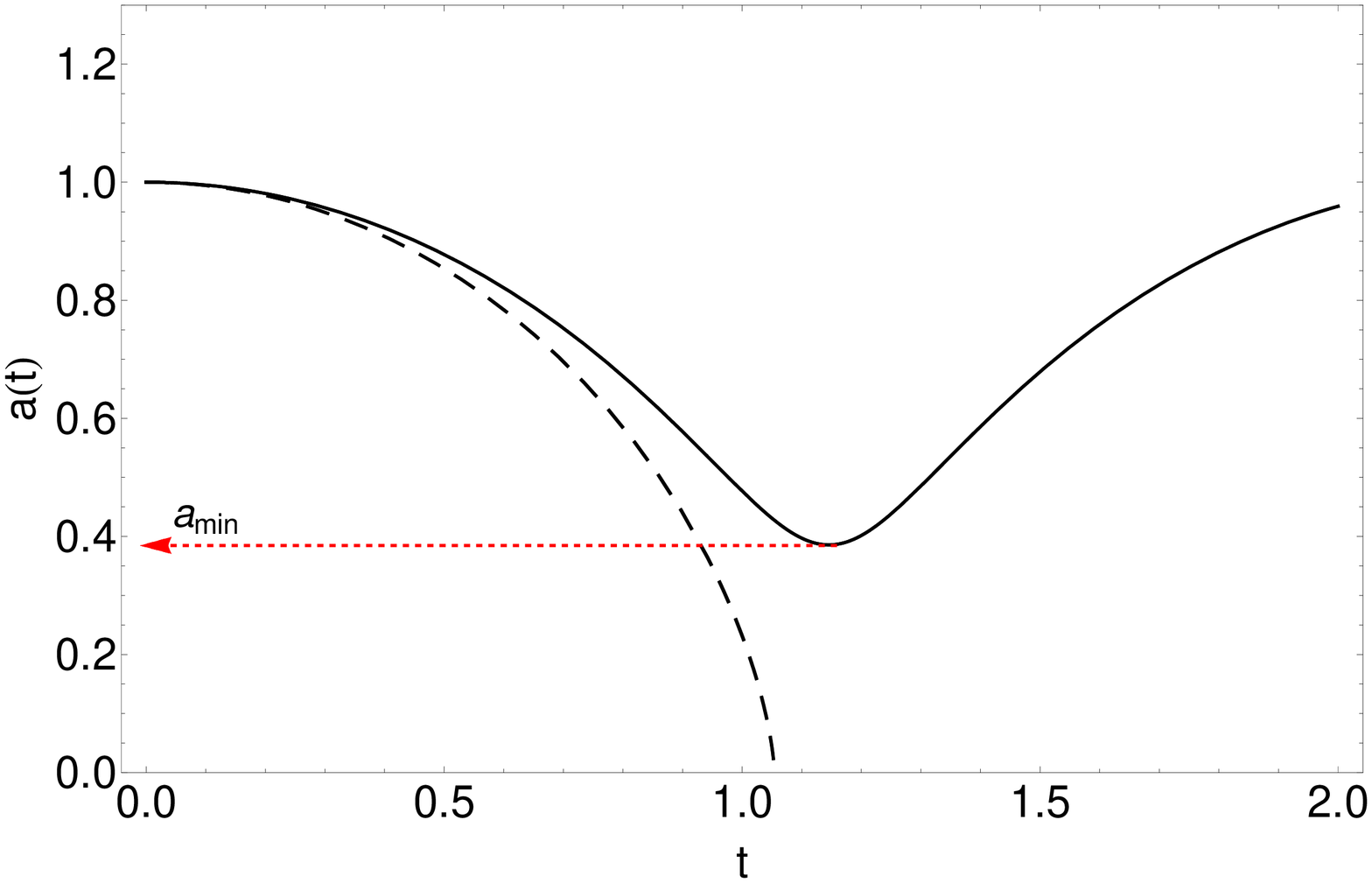}
\includegraphics[scale=0.335]{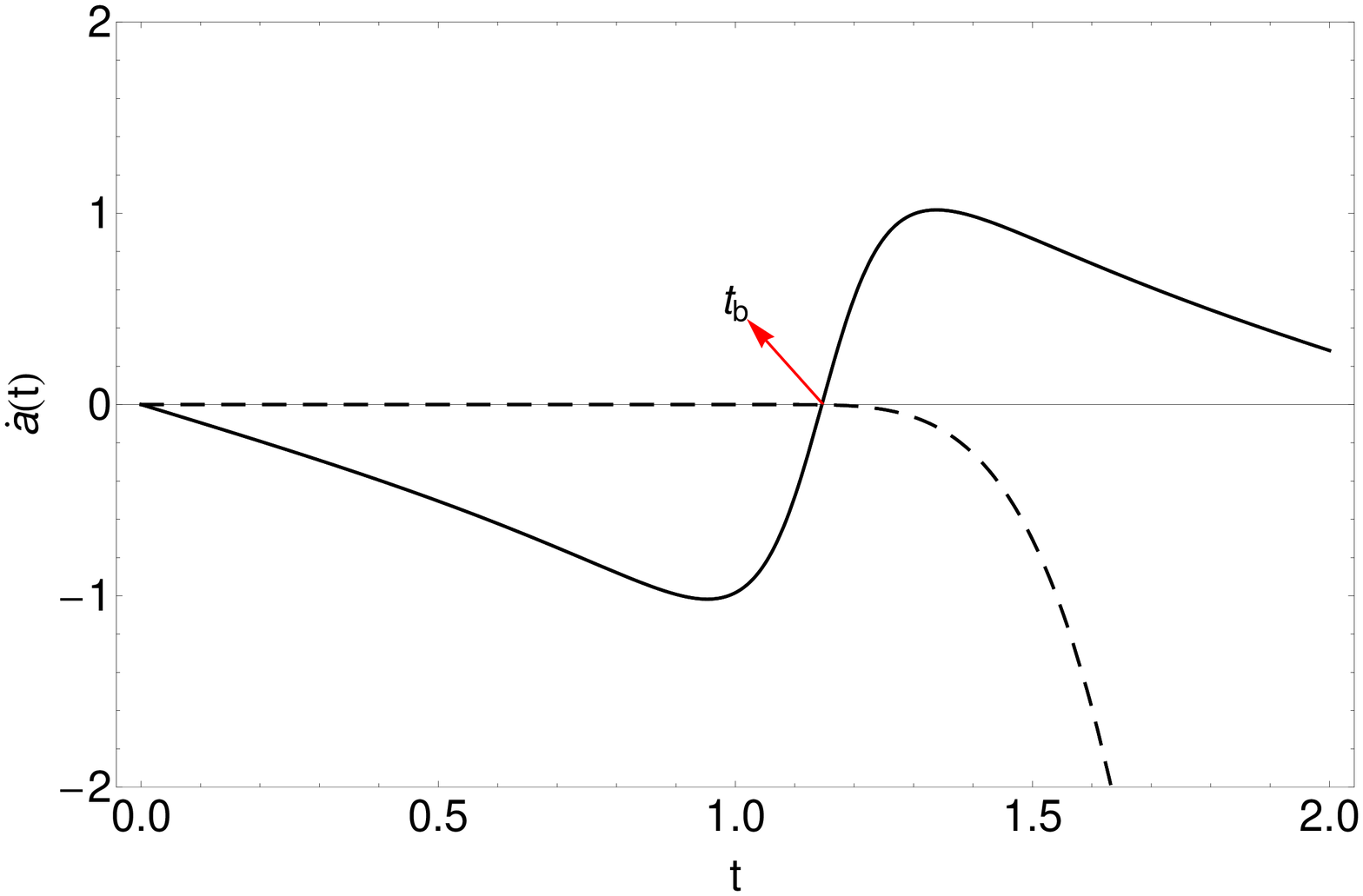}
\caption {Time behavior of the scale factor (left panel) and the collapse velocity (right panel) for $C=1.11$,  $a_i=1$ and $\dot{a}(t_i)=0$, $D=0.08$ (full curve) and $D=0.0$ (dashed curve).}\label{scf}
\end{center}
\end{figure}

In order to get a better understanding of the situation we perform a numerical simulation for the time behavior of the scale factor, the collapse velocity, its acceleration and Kretschmann scalar, by solving the second part of (\ref{FERE}) numerically and taking the first part as the initial constraint. The left panel of figure \ref{scf} shows that if the spin effects are neglected the collapse process terminates in a spacetime singularity (dashed curve) where the scale factor vanishes at a finite amount of comoving time. While, as the full curve shows, the scale factor begins its evolution from its initial value but deviates from the singular curve as the collapse advances. It then reaches a minimum value ($a_{{\rm min}}$) at the bounce time, $t=t_b$, after which the collapsing phase turns to an expansion. The scale factor never vanishes and hence the spacetime is regular throughout the contracting and expanding phases. 
\begin{figure}
\begin{center}
\includegraphics[scale=0.33]{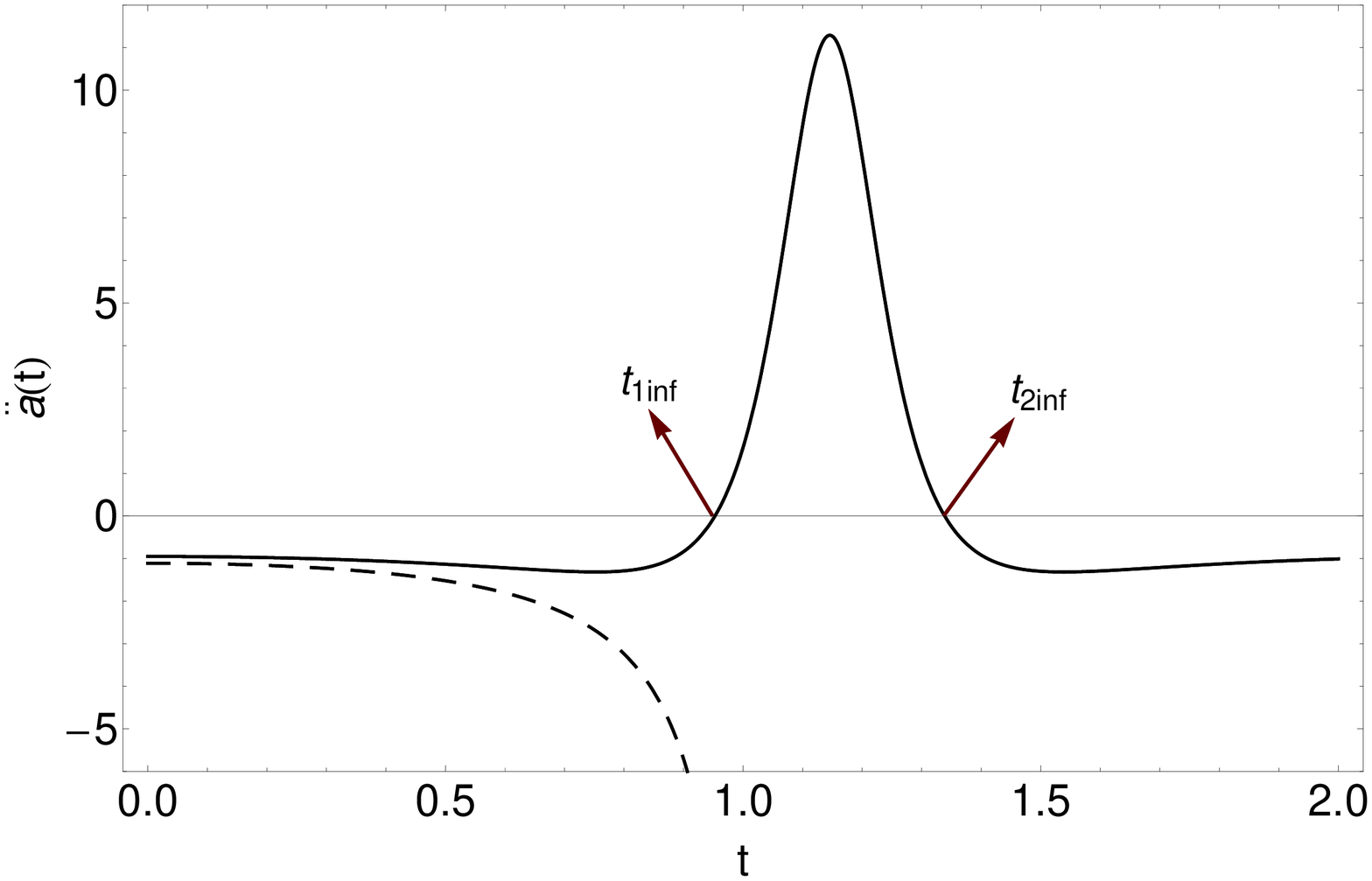}
\includegraphics[scale=0.34]{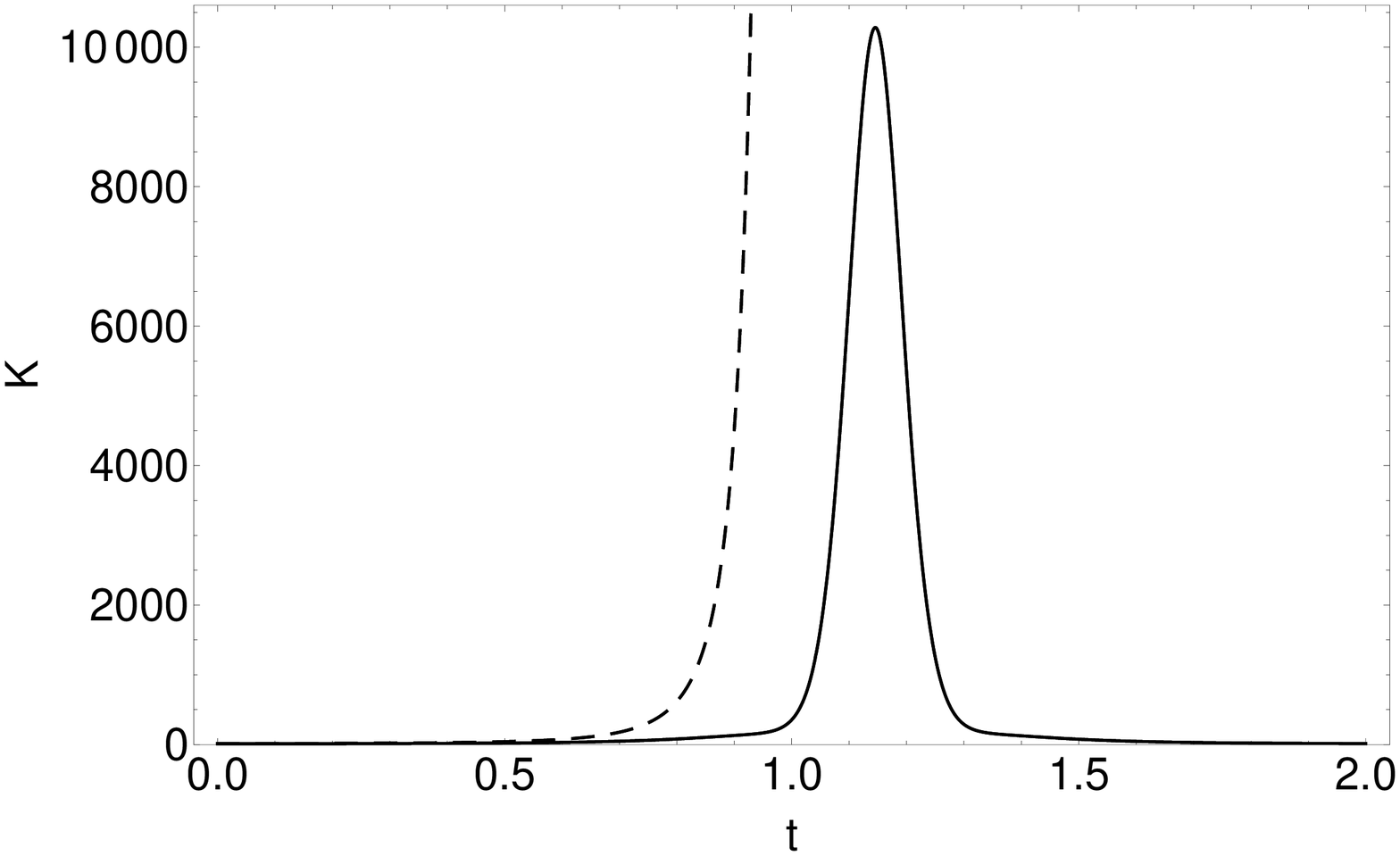}
\caption {Time behavior of collapse acceleration (left panel) and Kretschmann scalar (right panel) for $C=1.11$, , $a_i=1$ and $\dot{a}(t_i)=0$, $D=0.08$ (full curve) and $D=0.0$ (dashed curve).}\label{scf1}
\end{center}
\end{figure}
The diagram for the speed of collapse  indeed verifies such a behavior (see the full curve in the right panel of figure \ref{scf}) where the collapse begins at rest with the speed changing its sign from negative to positive values at the bounce time. The behavior of the collapse acceleration gives us more interesting results. We see that $\ddot a$ changes its sign at two inflection points (see the left panel of figure \ref{scf1}) in such a way that for $t<t_{{\rm 1inf}}$ the collapse undergoes an accelerated contracting phase ($\ddot a<0$ and $\dot a<0$). For the time interval $t_{{\rm 1inf}}<t<t_b$ the collapse process experiences a decelerated contracting phase where $\ddot a<0$ and $\dot a<0$. After the bounce occurs, the scenario enters an inflationary expanding regime till the second inflection point is reached, i.e., $\ddot a>0$ and $\dot a>0$ for $t_b<t<t_{{\rm 2inf}}$. Finally a decelerating expanding phase commences once the collapse acceleration passes through its second inflection point (the same cosmological scenario has been discussed in \cite{G1986}). The collapsing cloud then disperses at later times. Concomitantly, the Kretschmann scalar increases toward a maximum value but remains finite during the whole process of contraction and expansion (see the full curve in the right panel of figure \ref{scf1}) signaling the avoidance of spacetime singularity.
\par
\begin{figure}
\begin{center}
\includegraphics[scale=0.4]{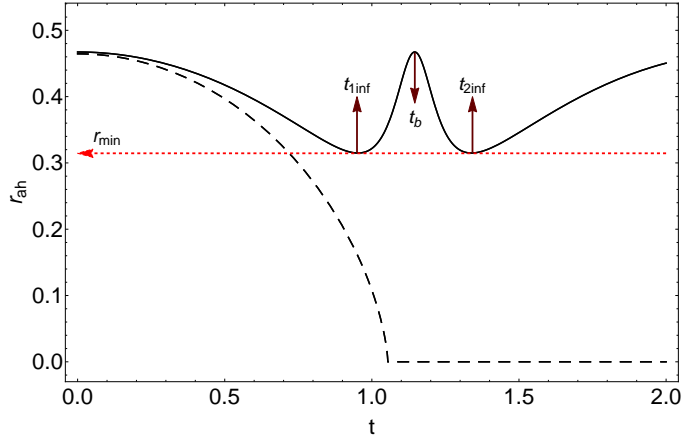}
\caption {Time behavior of the apparent horizon curve for $a_i=1$ and $\dot{a}(t_i)=0$, $C=1.11$, $D=0.08$ (full curve) and $C=1.11$, $D=0$ (dashed curve).}\label{scf2}
\end{center}
\end{figure}

 Now, what would happen to the formation of apparent horizon during the entire evolution of the collapsing cloud and specially whether the bounce is visible or not? In order to answer this question we proceed by recasting the metric \ref{FRWL} into the double-null form as
\be\label{DN}
ds^2=-2d\zeta^+d\zeta^{-}+R^2d\Omega^2,
\ee
with the null one-forms defined as
\bea\label{NOF}
d\zeta^+\!\!\!&=&\!\!\!-\f{1}{\sqrt{2}}\left[dt-\f{a}{\sqrt{1-kr^2}}dr\right],\nn
d\zeta^-\!\!\!&=&\!\!\!-\f{1}{\sqrt{2}}\left[dt+\f{a}{\sqrt{1-kr^2}}dr\right].
\eea
From the above expressions we can easily find the null vector fields as
\bea\label{NVF}
\partial_+=\f{\partial}{\partial\zeta^+}\!\!\!&=\!\!\!&-\sqrt{2}\left[\partial_t-\f{\sqrt{1-kr^2}}{a}\partial_r\right],\nn
\partial_-=\f{\partial}{\partial\zeta^-}\!\!\!&=\!\!\!&-\sqrt{2}\left[\partial_t+\f{\sqrt{1-kr^2}}{a}\partial_r\right].
\eea
The condition for the radial null geodesics, $ds^2=0$, leaves us with the two kinds of null geodesics characterized by $\zeta^+=constant$ and $\zeta^-=constant$. The expansions along these two congruences are given by
\be\label{exp}
\theta_{\pm}=\f{2}{R}\partial_{\pm}R.
\ee
In a spherically symmetric spacetime, the Misner-Sharp quasi-local mass which is the total mass within the radial coordinate $r$ at the time $t$ is defined as \cite{MSMASS}
\bea\label{MSM}
m(t,r)&=&\f{R(t,r)}{2}\left(1+g^{\mu\nu}\partial_{\mu}R(t,r)\partial_{\nu}R(t,r)\right)\nn
&=&\f{R(t,r)}{2}\left(1+\f{R(t,r)^2}{2}\theta_+\theta_-\right).
\eea
Therefore, it is the ratio $2m(t,r)/R(t,r)=\dot{R}(t,r)^2+kr^2$ that controls the formation or otherwise of trapped surfaces so that the apparent horizon defined as the outermost boundary of the trapped surfaces is given by the condition $\theta_{+}\theta_{-}=0$ or equivalently $2m(t,r_{{\rm ah}}(t))=R(t,r_{{\rm ah}}(t))$. The equation for the apparent horizon curve then reads
\be\label{AH}
R(t,r_{{\rm ah}}(t))^{-2}=\left(\f{\dot a(t)}{a(t)}\right)^2+\f{k}{a(t)^2},
\ee
or by the virtue of the first part of \ref{FERE}
\be\label{rah}
r_{{\rm ah}}(a(t))=\left[\f{2C}{a(t)}-\f{D}{a(t)^{4}}\right]^{-\f{1}{2}},
\ee
which shows the time at which the shell labeled by $r$ becomes trapped. Figure \ref{scf2} shows the time behavior of the apparent horizon where we see that in the presence of spin effects, the apparent horizon decreases (see the full curve) for a while to a minimum value where the first inflection point is reached ($\ddot{a}(t_{{\rm 1inf}})=0$). It then increases to a finite maximum at the bounce time and converges again in the post-bounce regime to the same minimum value where the acceleration vanishes for the second time. The apparent horizon goes to infinity at later times. In order to find the minimum value for the apparent horizon curve we can easily extremize equation (\ref{AH}) to get 
\be\label{EXTRAH}
a_{\star}=\left(\f{2D}{C}\right)^{\f{1}{{3}}}.
\ee
Therefore there exists a minimum radius
\be\label{MINr}
r_{{\rm min}}=r_{{\rm AH}}(a_{\star})=\f{1}{\sqrt{2}a_i}\left(\f{\hbar}{\pi G\rho_i A_0}\right)^{\f{1}{3}},
\ee
for which if the boundary of the collapsing object is taken as $r_b<r_{{\rm min}}$, the apparent horizon does not develop throughout the collapsing and expanding phases and therefore the bounce is uncovered. From the viewpoint of equation (\ref{AH}), we can also deduce that since the collapse velocity is bounded the apparent horizon never converges to zero implying that there exists a minimum radius (the dashed red curve) below which no horizon would form to meet the boundary of the collapsing object. However, the apparent horizon does not diverge at the bounce since $k>0$ and to which extent it could grow depends on the initial configuration of the stellar object. When the spin effects are absent, the apparent horizon decreases monotonically (see the dashed curve) to finally cover the resulted singularity. There can not be found any minimum for the boundary of the collapsing cloud so that the formation of the horizon can be prevented.
The existence of a minimum value for the boundary implies that there exists a minimum value for the total mass contained within the collapsing cloud. Let us be more precise. Using equation (\ref{MSM}) we can re-write the dynamical interior field equations (\ref{FERE}) as
\bea\label{REFES12}
\partial_r m(t,r)&=&4\pi G\rho_{{\rm eff}}R(t,r)^2\partial_r R(t,r),\nn
\partial_t m(t,r)&=&-4\pi Gp_{{\rm eff}}R(t,r)^2\partial_t R(t,r),
\eea
whereby integration of the first part gives
\be\label{mtr}
m(t,r)=\f{4\pi G}{3}r^3\rho_i a_i^3\left[1-\f{\pi G \hbar^2}{4A_0^2}\rho_i \left(\f{a_i}{a(t)}\right)^3\right].
\ee
The above expression together with (\ref{EXTRAH}) and (\ref{MINr}) gives the threshold mass confined within the radius $r_{{\rm min}}$ as
\be\label{THMA}
m_{\star}=m(a_{\star},r_{{\rm min}})=\f{\hbar}{\sqrt{2}A_0}.
\ee
Thus, if the total mass is chosen so that $m<m_{\star}$, there would not exist enough mass within the collapsing cloud at the later stages of the collapse to get the light trapped and as a result, formation of the apparent horizon is avoided. 

 Furthermore, the time derivative of the mass function, using the second part of (\ref{REFES12})
\be\label{TDMASS}
\partial_t m(t,r)=\f{\pi^2 G^2 \hbar^2\rho_i^2 a_i^6}{A_0^2a(t)^4}r^3\dot{a}(t),
\ee
is negative throughout the contracting phase. This may be interpreted as if some mass may be thrown away from the stellar object till the bounce time is approached. At this time, $m(a_{{\rm min}},r)=0$ which can be imagined as the whole evaporation of the collapsing cloud. After this time, when the expanding phase begins, the ejected mass may be regained since $\dot{m}|_{(t>t_b)}>0$. We note that such a behavior is due to the homogeneity of the model since all the shells of matter collapse or expand simultaneously. For the case of dust fluid considered here, the exterior region of the star can be modeled by Schwarzchild spacetime since the spin effects are negligible at the early stages of the collapse. However, as the collapse advances, the mass profile is no longer constant due to the presence of negative pressure which originates from the spin contribution. Hence, at the very late stages of the collapse, The Schwarzchild spacetime may not be a suitable candidate for the matching process and instead the interior region should be smoothly matched to the exterior generalized Vaidya metric \cite{Santos-1985}. Let us consider the exterior spacetime in retarded null coordinates as
\be \label{EXO}
ds_{{\rm out}}^2=f(u,r_v)du^2+2dudr_v-r_v^2(d\theta^2+\sin^2\theta d\phi^2),
\ee
where $f(u,r_v)=1-2{\rm M}(r_v,u)/r_v$ with ${\rm M}(r_v,u)$ being the Vaidya mass. We label the exterior coordinates as $\left\{X_{{\rm out}}^{\mu}\right\}\equiv\left\{u,r_v,\theta,\phi\right\}$ where $u$ is the retarded null coordinate labeling different shells of radiation and $r_v$ is the Vaidya radius. The above metric is to be matched through the timelike hypersurface $\Sigma$ given by the condition $r=r_b$ to the interior line element given by (\ref{FRWL}). The interior coordinates are labeled as $\left\{X_{{\rm in}}^{\mu}\right\}\equiv\left\{t,r,\theta,\phi\right\}$.  The induced metrics from the interior and exterior spacetimes close to $\Sigma$ then read
\be\label{INMI}
ds_{\Sigma{\rm out}}^2=dt^2-a^2(t)r_{b}^2(d\theta^2+\sin^2\theta d\phi^2),
\ee
and
\be \label{INMEX}
ds_{\Sigma{{\rm out}}}^2=\left[f\big(u(t),r_v\big)\dot{u}^2+2\dot{r}_v\dot{u}\right]dt^2-r_v^2(t)(d\theta^2+\sin^2\theta d\phi^2).
\ee
Matching the induced metrics give
\be\label{MINMET}
f\dot{u}^2+2\dot{r}_v\dot{u}=1,~~~~r_v(t)=r_ba(t).
\ee
The unit vector fields normal to the interior and exterior hypersurfaces can be obtained as
\bea\label{UNVF}
n^{{\rm in}}_{\mu}&=&\left[0,\f{a(t)}{\sqrt{1-kr^2}},0,0\right],\nn
 n^{{\rm out}}_{\mu}&=&\frac{1}{\left[f(u,r_v)\dot{v}^2+2\dot{r}_v\dot{u}\right]^{\frac{1}{2}}}\left[-\dot{r}_v,\dot{u},0,0\right].
\eea
The extrinsic curvature tensors for the interior and exterior spacetimes are given by
\be\label{EXCIN}
K^{{\rm in}}_{ab}=-n_{\mu}^{{\rm in}}\left[\frac{\partial^2X_{{\rm in}}^{\mu}}{\partial y^a\partial y^b}+\hat{\Gamma}^{\mu\,{\rm in}}_{\,\nu\sigma}\frac{\partial X_{{\rm in}}^{\nu}}{\partial y^a}\f{\partial X_{{\rm in}}^{\sigma}}{\partial y^b}\right],
\ee
and 
\be\label{EXCOUT}
K^{{\rm out}}_{ab}=-n_{\mu}^{{\rm out}}\left[\frac{\partial^2X_{{\rm out}}^{\mu}}{\partial y^a\partial y^b}+\big\{^{\,\mu}_{\nu\sigma}\big\}^{\rm out}\frac{\partial X_{{\rm out}}^{\nu}}{\partial y^a}\f{\partial X_{{\rm out}}^{\sigma}}{\partial y^b}\right],
\ee
respectively where $y^a=\{t,\theta,\phi\}$ are coordinates on the boundary. We note that in computing the components of the extrinsic curvature of  the interior spacetime, the general affine connection should by utilized. However, from equations (\ref{AFC}-\ref{FEEC}) together with (\ref{FC}), we see that the affine connection is finally obtained linearly with respect to the spin density tensor. Therefore, by a suitable spacetime averaging, only the Christoffel symbols would remain to be used in (\ref{EXCIN}). The non-vanishing components of the extrinsic curvature tensors then read
\bea\label{EXCCS}
K_{tt}^{{\rm in}}&=&0,~~~~~~K^{\!\theta\,{\rm in}}_{\,\,\theta}=K^{\!\phi\,{\rm in}}_{\,\,\phi}=\frac{\sqrt{1-kr_b^2}}{r_ba(t)},\nn
K_{tt}^{{\rm out}}&=&-\frac{\dot{u}^2\left[ff_{,r_v}\dot{u}+f_{,u}\dot{u}+3f_{,r_v}\dot{r}_v\right]
+2\left(\dot{u}\ddot{r}_v-\dot{r}_v\ddot{u}\right)}{2\left(f\dot{u}^2+2\dot{r}_v\dot{u}\right)^{\frac{3}{2}}},\nonumber\\
K^{\!\theta\,{\rm out}}_{\,\,\theta}&=&K^{\!\theta\,{\rm out}}_{\,\,\theta}=\frac{f\dot{u}+\dot{r}_v}{r_v\sqrt{f\dot{u}^2+2\dot{r}_v\dot{u}}}.
\eea
Matching the components of extrinsic curvatures on the boundary give
\bea
&&f\dot{u}+\dot{r}_v=\sqrt{1-kr_b^2},\label{matchinout1}\\
&&\dot{u}^2\left[ff_{,r_v}\dot{u}+f_{,u}\dot{u}+3f_{,r_v}\dot{r}_v\right]
+2\left(\dot{u}\ddot{r}_v-\dot{r}_v\ddot{u}\right)=0.\label{matchinout2}\nn
\eea
A straightforward but lengthy calculation reveals that (\ref{matchinout2}) results in $f(r_v,u)=f(r_v)$ on the boundary \cite{Hussain-PRD-2011}. Furthermore from (\ref{matchinout1}) and the first part of (\ref{MINMET}) we get
\be\label{rvdot}
\dot{r}_v=-(1-f-kr_b^2)^{\f{1}{2}},
\ee
whence using the second part of (\ref{MINMET}) we readily arrive at the following equality
\be\label{EQMM}
{\rm M}(r_v)=m(t,r_b).
\ee
Thus from the exterior viewpoint, equation (\ref{THMA}) implies that there can be found a mass threshold so that for the mass distributions below such a threshold, the apparent horizon would fail to intersect the surface boundary of the collapsing cloud. Moreover, in view of (\ref{mtr}), we observe that as the scale factor increases in the post-bounce regime, the second term decreases and vanishes at late times. This leaves us with a Schwarzchild exterior spacetime with a constant mass. 
\section{Concluding remarks}
\label{con}

We studied the process of gravitational collapse of a massive star whose matter content is a homogeneous Weyssenhoff fluid in the context of EC theory. Such a fluid is considered as a perfect fluid with spin correction terms that come from the presence of intrinsic angular momentum of fermionic particles within a real star. The main objective of this paper was to show that, contrary to the OS model, if the spin contributions of the matter sources are included in the gravitational field equations, the collapse scenario does not necessarily end in a spacetime singularity. The spin effects can be negligible at the early stages of the collapse, while as the collapse proceeds, these effects would play a significant role in the final  fate of the collapse scenario. This situation can be compared to the singularity removal for a FLRW spacetime in the very early universe, as we go {\it backwards-in-time} \cite{spin-bounce,POPPRD}. We showed that in contrast to the homogeneous dust collapse which leads inevitably to the formation of a spacetime singularity, the occurrence of such a
singularity is avoided and instead a bounce occurs at the end of the
contracting phase. The whole evolution of the star experiences four
phases that the two of which are in the contracting regime and the other two ones are in the
post-bounce regime. While, in the homogeneous dust case without spin
correction terms, the singularity is necessarily dressed by an event
horizon, formation of such a horizon can be always prevented by
suitably choosing the surface boundary of the collapsing star. This signals that there exists a critical threshold value for the mass content, below which no horizon would form. The same picture can be found in \cite{BMMM} where the non-minimal coupling of gravity to fermions is allowed. Besides the model presented here, non-singular scenarios have been reported in the literature within various models such as $f(R)$ theories of gravity in Palatini \cite{fr} and metric \cite{npcb} formalisms, non-singular cosmological settings in the presence of spinning fluid in the context of EC theory \cite{bretchet}, bouncing scenarios in brane models \cite{brb} and modified Gauss-Bonnet gravity \cite{mgbb} (see also \cite{repb} for recent review). While the spacetime singularities could generically occur as the end-product of a continual gravitational collapse, it is widely believed that in the very final stages of the collapse where the scales are comparable to Planck length and extreme gravity regions are dominant,  quantum corrections could generate a strong negative pressure in the interior of the cloud to finally resolve the classical singularity \cite{QGRS}. Finally, as we near to close this paper it deserves to point out that, quantum effects due to particle creation could possibly avoid the cosmological \cite{pfprd} as well as astrophysical singularities \cite{BACK}.

\section{Acknowledgments}
The authors would like to sincerely thank the anonymous referee for constructive and helpful comments to improve the original manuscript.

\end{document}